# Killingbeck Mass Spectrum of Heavy Quarkonia Under the Influence of External Magnetic and AB Flux Fields

M. Eshghi[1,*], H. Mehraban[2], S.M. Ikhdair[3,4]

[1] *Young Researchers and Elite club, Central Tehran Branch, Islamic Azad University, Tehran, Iran*

[2] *Faculty of Physics, Semnan University, Semnan, Iran*

[3] *Department of Physics, Faculty of Science, An-Najah National University, Nablus, West Bank, Palestine*

[4] *Department of Electrical Engineering, Near East University, Nicosia, Northern Cyprus, Mersin 10, Turkey*

## Abstract

We present a more general form of the Schrödinger equation in curvilinear space that is exposed to external fields. Thereby, we solve the wave equation with the Cornell plus harmonic potentials (CHP) under the influence of the magnetic and Aharonov-Bohm (AB) flux fields using curvilinear coordinates system in such space. With this requirement, the energy spectrum and the corresponding wave functions are calculated by means of the series method. The mass spectrum of both charmonium and bottomonium states consisting of quark and antiquark are calculated. In addition, the main thermodynamic functions such as the free energy, the mean energy, the entropy, the specific heat, the persistent currents and magnetization are obtained by using the characteristic function. We draw the resulting energy states and the thermodynamic quantities versus some potential parameters, magnetic field and temperature to see their numerical behaviors. Finally, we give some discussions to our results.

**Keywords:** Schrödinger equation; thermodynamic functions; series method; partition function; Quark and antiquark.

**PACS numbers:** 03.65.Ge; 03.65.Db; 05.70.Ce; 02.30.Gp.

## 1. Introduction

To study any physical model, we need to perform calculations for the essential physical quantities as the initial stage. Hence, in searching for more accurate solutions of the non-relativistic and relativistic wave equations has turned to become a principal part from the beginning of the quantum, atomic and nuclear physics [1-12].

---

[*] ***Email corresponding author:*** *eshgi54@gmail.com; m.eshghi@semnan.ac.ir; meshghi@ihu.ac.ir*

Furthermore, the investigation of the non-relativistic and relativistic quantum dynamics of charged particles in the presence of vector and scalar fields has also become very important. In other words, the study of the quantum dynamics of these charged particles in presence of external magnetic and Aharonov-Bohm (AB) flux fields [13], which are perpendicular to the plane where the particles are confined, has been carried out over the past years. In fact, the research of systems of non-relativistic and relativistic charged particles that are confined to the magnetic fields has attracted much attention due to the possible applications.

For example, some of applications were used in cosmic string [14], semiconductor structures [15], biology [16], molecular vibrational and rotational spectroscopy of molecular physics [17], chemical physics [18], and other scientific areas such as environmental sciences [19], and graphene [11-13, 20]. Also, the structures of the charged carriers are restricted mainly by the quantum potentials. These confinements may result in the formation of discrete energy levels, drastic changes of optical absorption spectra and so forth [21-27].

In quantum physics, the non-relativistic equation describing charged particles moving in a whole range of configurations is reducing to a biconfluent Heun equation. In this case, we have such examples as: charged particles moving under a magnetic field and Coulomb repulsive potential or in a homogeneous magnetic field and a two-dimensional parabolic potential. Also, if one studies only the linear systems, then it would result in different forms of the confluent hypergeometric equations. However, there are examples such as gravitational physics, graphene and other areas in physics where the Heun equation appears in the nonlinear systems [28-30]. Occasionally, choosing the special potential model of the physical system is a suitable reason for the appearance of Heun function. In this regard, there are a large number of potentials of physical importance used in the Schrödinger, Dirac and Dirac-Weyl equations that may be transformed into the Heun biconfluence function (BHE) [12, 31-36].

Further, in studying the behavior of a single charged particle under the magnetic fields, we can suppose the quantum systems as a Fermi gas. In this case, we need to consider the behavior of the collective interactions of the charged particles in such systems [26, 37]. Therefore, the aim is to determine the behavior of the thermodynamic quantities and for a further understanding to the thermal effects on various physical quantities. So we need to investigate the energy behavior of the Schrodinger equation for a particle in a gas system with an interaction potential and also the main features of the particle with respect to the thermodynamic properties. Recently, many authors have studied the thermodynamic properties of many potential model systems [38-41].

Therefore, our aim is to study the Cornell plus harmonic oscillator potential [33, 42] having the form:

$$V(r) = ar^2 + br - \frac{g}{r}, \tag{1a}$$

where $a$, $b$ and $g$ are the potential parameters. This potential model has no analytical solutions except for quasi-exactly solutions. The condition on the potential parameters are determined as

$$g = (n + l + 1)b\sqrt{1/2a\mu}, \tag{1b}$$

where $\mu = m_c/2$ and $\mu = m_b/2$ for charmonium and bottomonium consisting of quark and antiquark of same flavor, respectively.

The fore mentioned potential, $V = V(a,b,g;r)$, is a dependent on different parameters and that the changing of these parameters is effective and hence leading to the changing the value of calculated energy and other quantities dependent on the energy as well. In fact, the Cornell plus harmonic confining potential is mostly used to study quantum dots [43] and also heavy quarkonia [44, 45]. Hence, for a further study, we refer to Refs.[46, 47]. The potential used for obtaining the mass spectrum of heavy mesons consisting of heavy quarks and antiquarks $(\Psi(b\bar{b}), \psi(c\bar{c}))$, is called the Cornell type, i.e., Coulomb plus linear terms. This type of interaction potential model is accompaniment by lattice quantum charomodynamics calculations [48]. The Coulomb term is to be liable for the interaction at small distances and linear term leads to the quark confinement. In region between, the quark-antiquark interaction can also be studied using Coulomb plus power potential [49].

The main task of the present paper is to solve the non-relativistic equation with the Cornell plus harmonic potential model in the presence of external magnetic and *AB* flux fields. It is worthy to note that we introduce a general form of the Schrodinger equation in a curvilinear space. Namely, we use the curvilinear coordinates system in such space since the form and shape of every equation is covariant under every coordinate transformation $\left(\bar{x}^i = x^{-i}(x^1,...,x^N), i = 1,...,N\right)$. In other words, mathematical form and shape of the equation does not change in the curvilinear coordinate systems. Therefore, in this case, we can simply work with these coordinate systems by using the simple transformation in the curved space.

Hence, we can obtain the energy spectrum and the corresponding wave functions by applying the series method. Then, we calculate the main thermodynamic functions using the resulting energy spectrum. We also discuss about the obtained analytical results.

The structure of this paper is as follows. In Section 2, we present the method used to solve the Cornell-plus-harmonic potential model under the external magnetic and AB flux fields and obtain the energy and wave functions. Also, we calculate the mass spectrum of heavy quarkonia using the resulting bound state energy levels. In Section 3, we calculate the thermodynamic functions from the partition function. Finally, Sections 4 and 5 are devoted for our present results and discussions.

## 2. Theory and Calculations

Consider a single particle of mass moving with the energy $E_n$ in the three-dimensional curvilinear coordinates made by $(x^1, x^2, x^3)$ in a Rimanian manifold $M = \mathfrak{R}^3$ [50].

The Schrödinger equation for a non-relativistic particle can be written as

$$\left(-\frac{\hbar^2}{2\mu}\nabla^2 + V(\vec{X}) - E_n\right)\Psi(\vec{X}) = 0, \tag{2}$$

where $\mu$ is the effective mass, $\Psi(\vec{X})$ and $E_n$ are the wave functions and the eigenspectra of the quantum system, respectively.

Using the transformation given by $\left(-i\hbar\vec{\nabla} \to -i\hbar\vec{\nabla} + e\vec{A}/c\right)^2$, with $\vec{A}$ is the vector potential, into Eq. (2), one obtains

$$\left(\nabla^2 + \frac{e}{\hbar c}\vec{A}\cdot\vec{\nabla} - \left(\frac{e}{\hbar c}\right)^2 \vec{A}^2 - \frac{2\mu}{\hbar^2}\left[V(\vec{X}) - E_n\right]\right)\Psi(\vec{X}) = 0. \tag{3}$$

where $\vec{\nabla}\cdot\vec{A} = 0$.

Now, we can write first term of Eq. (3) as follows [34, 51]:

$$\nabla^2 = \Gamma_{ul}^u \partial^l + g^{kl}\partial_l \partial_k = g^{-1/2}\partial_l\left(g^{1/2}\partial^l\right), \tag{4}$$

where $\partial^l \equiv \partial/\partial x_l$ and $\Gamma^n{}_{kl}$ are Christoffel symbol of the second kind and being defined as $\Gamma^s{}_{kl} = \frac{1}{2} g^{su} \left( \frac{\partial g_{sk}}{\partial x^l} + \frac{\partial g_{sl}}{\partial x^k} - \frac{\partial g_{kl}}{\partial x^u} \right)$, with $s,k,u = 1,2,...,N$, where $g^{su}$ and $g_{sl}$ are the contravariant and covariant components of the metric tensor, respectively. The second and the third terms of Eq. (3) reads as

$$\vec{A} \cdot \vec{\nabla} = A_k \partial^k, \tag{5}$$

and

$$A^2 = g^{ik} A_i A_k = A_i^2. \tag{6}$$

Upon substituting Eqs. (4), (5) and (6) into Eq. (3), we obtain

$$\left( g^{-1/2} \partial_l \left( g^{1/2} \partial^l \right) + i \frac{e}{\hbar c} A_k \partial^k - \left( \frac{e}{\hbar c} \right)^2 A_i^2 - \frac{2\mu}{\hbar^2} [V(\vec{X}) - E_n] \right) \Psi(\vec{X}) = 0. \tag{7}$$

where Eq. (7) is the Schrödinger equation in a Rimanian manifold $N$-dimensional $M = \Re^3$ for every point such as $p \in M$.

Therefore, the Schrodinger equation is obtained in the two-dimensional curvilinear coordinates system for $\forall \ p \in M$, plane polar coordinates system, as below

$$\left\{ \frac{1}{h_1 h_2} \partial_l \left( \frac{h_1 h_2}{h_l^2} \partial^l \right) + i \frac{e}{\hbar c} A_k \partial^k - \left( \frac{e}{\hbar c} \right)^2 A_i^2 - \frac{2m}{\hbar^2} [E_n - V(\vec{X})] \right\} \Psi(\vec{X}) = 0, \tag{8}$$

where, we supposed the Lame coefficients of this system as $h_1, h_2$ and used $h_k = \sqrt{g_{kk}}$ with $k = 1,2$ and $g = \det(g_{jk}) = h_1 h_2$ inserted into Eq. (8) with Lamme coefficients taken as $h_1 = 1$, $h_2 = r$. Here, we assumed that the curvilinear coordinates systems be as orthogonal.

In order to calculate the energy levels and wave functions under the external magnetic and AB flux fields can be essentially performed with a few algebraic calculations and using the two-dimensional wave functions:

$$\psi(r, \phi) = \frac{1}{\sqrt{2\pi}} e^{im\phi} u(r), \tag{9}$$

where $m$ is the magnetic quantum number. We can simply obtain the radial part of the Schrödinger equation in polar coordinates $(r, \phi)$, Eq. (8), with the component of the vector potential in a simple form: $A = (0, Br/2 + \Phi_{AB}/2\pi r, 0)$ [33, 34] and setting $u(r) = f(r)/\sqrt{r}$ as below

$$\left[\frac{d^2}{dr^2} - \frac{2\mu E}{\hbar^2} + \left(m + \frac{\Phi_{AB}}{\Phi_0}\right)\frac{\mu\omega_c}{\hbar} - \frac{\left(m + \frac{\Phi_{AB}}{\Phi_0}\right)^2 - \frac{1}{4}}{r^2}\right. \tag{10}$$

$$\left. + \frac{2\mu g}{\hbar^2}\frac{1}{r} + \frac{2\mu b}{\hbar^2}r - \left(\frac{2\mu a}{\hbar^2} + \left(\frac{\mu\omega_c}{2\hbar}\right)^2\right)r^2\right]f(r) = 0,$$

where $\Phi_0 = hc/e$ and the cyclotron frequency is as $\omega_c = eB/\mu c$.

The above equation is the radial biconfluent Henu's equation that takes on the Schrödinger form. Now making further change of variables $\chi = \gamma^{1/2}r$, we can find

$$\left[\frac{d^2}{d\chi^2} + \frac{\lambda_{E,B,\Phi}}{\gamma} - \frac{\alpha^2 - \frac{1}{4}}{\chi^2} + \frac{\eta}{\chi} + \tilde{b}\chi - \chi^2\right]f_{nm}(\chi) = 0, \tag{11}$$

where we have used the assignments $\alpha^2 = \left(m + \frac{\Phi_{AB}}{\Phi_0}\right)^2$, $\gamma^2 = \frac{2\mu a}{\hbar^2} + \left(\frac{\mu\omega_c}{2\hbar}\right)^2$

$\tilde{b} = -\frac{2\mu b}{\gamma^{3/2}\hbar^2}$, $\eta = \frac{2\mu g}{\gamma^{1/2}\hbar^2}$ and $\lambda_{E,B,\Phi} = \frac{2\mu E}{\hbar^2} - \left(m + \frac{\Phi_{AB}}{\Phi_0}\right)\frac{\mu\omega_c}{\hbar}$.

Thus, Eq. (11) is in the form of Eq. (1.2.5) given by Ref. [36].

The differential Eq. (11) has two singular points: an irregular one at infinity and a regular one at the origin. In order to obtain the energy spectrum of Eq. (11), it is convenient to analyze its asymptotic behavior. Unfortunately, in the general case, there is no closed expression for large values of argument for the asymptotic behavior of the biconfluent Heun function. In order to obtain a polynomial form with the conditions on parameters of the biconfluent Heun function, can only be calculated by analyzing each coefficient separately.

In discussing the asymptotic behavior, the solution of Eq. (11) for small values of the variable $\chi$, at $f(0) = 0$, determined by centrifugal term and the asymptotic behavior of the solution of Eq. (10) for large values of the variable $\chi$, at $f(\chi \to \infty) \to 0$, determined by the oscillator terms. In this regard, the case when $\tilde{b} = \gamma = 0$ has been analyzed in [52]. After having made this analysis, now we can select an appropriate transformation for the function $f(\chi)$ as follows: $f_{nm}(\chi) = \chi^{\alpha-1/2}\exp\left[-(\chi^2 - \chi\tilde{b})/2\right]F(\chi)$. Therefore, Eq. (11), turns to become

$$\chi \frac{d^2 F(\chi)}{d\chi^2} + \left[ 2\left(\alpha + \frac{1}{2}\right) + \tilde{b}\chi - 2\chi^2 \right] \frac{dF(\chi)}{d\chi}$$
$$+ \left[ \eta + \tilde{b}\left(\alpha + \frac{1}{2}\right) + \left(\frac{\lambda_{E,B,\Phi}}{\gamma} + \frac{\tilde{b}^2}{4} - 2\left(\alpha + \frac{1}{2}\right) - 1\right)\chi \right] F(\chi) = 0. \tag{12}$$

Thus, Eq. (12) resembles the biconfluent Heun's (BCH) differential equation [31],

$$\frac{d^2 u}{d\zeta^2} + \frac{1}{\zeta}\left[\alpha + 1 - \beta\zeta - 2\zeta^2\right]\frac{du}{d\zeta} + \left[\frac{\delta + \beta + \alpha\beta}{2} + (\gamma - \alpha - 2)\zeta\right]\frac{u}{\zeta} = 0, \tag{13}$$

with the Heun's wave functions solution given by $u = H_B(\alpha, \beta, \gamma, \delta, -\zeta)$.

In comparing Eq. (12) with its counterpart Eq. (13), we can conclude that Eq. (12) is simply the BCH differential equation [31], whose solution is BCH function, $H_B$:

$$f_{nm}(\chi) = \chi^{\alpha+1/2} \exp\left[-\frac{\chi[\chi - \tilde{b}]}{2}\right]$$
$$\times H_B\left(2\left|m + \frac{\Phi_{AB}}{\Phi_0}\right|, -\frac{2\mu b}{\gamma^{3/2}\hbar^2}\right., \tag{14}$$
$$\left., \frac{-\frac{2\mu E}{\hbar^2} + \left(m + \frac{\Phi_{AB}}{\Phi_0}\right)\frac{\mu\omega_c}{\hbar}}{\gamma} + \frac{\mu^2 b^2}{\gamma^3 \hbar^4}, \frac{4\mu g}{\gamma^{1/2}\hbar^2}, -\chi\right).$$

Let us follow the results given in Eq. (12). We firstly define the parameters $P, Q, R$, and the function $\tilde{F}$ representing $F$ as follows

$$P = \left(\alpha + \frac{1}{2}\right), \quad Q = 1 - \left(\alpha + \frac{1}{2}\right)\frac{\tilde{b}}{\eta}, \quad \text{if} \quad \tilde{F} = F,$$
$$R = \frac{\lambda_{E,B,\Phi}}{\gamma^{1/2}} + \frac{\tilde{b}^2}{4} - 2\left(\alpha + \frac{1}{2}\right) - 1, \quad \text{if} \quad \tilde{F} = F, \tag{15}$$

with these parameters identifications, Eq. (12) can be casted

$$\chi \frac{d^2 \tilde{F}(\chi)}{d\chi^2} + \left[2P + \tilde{b}\chi - 2\chi^2\right]\frac{d\tilde{F}(\chi)}{d\chi} + (R\chi - Q\eta)\tilde{F}(\chi) = 0. \tag{16}$$

In order to solve the differential equation, we assume $\tilde{F}(\chi)$ is to be in the Frobenius series form as $\tilde{F}(\chi) = \sum_n C_n \chi^n$. So we substitute it into Eq. (16) and use the Frobenius serious method to obtain the following recurrence relation

$$C_{n+2} = \frac{\left[Q\eta - \tilde{b}(n+1)\right]C_{n+1} - (R-2n)C_n}{(n+2)(n+2P+1)}. \tag{17}$$

Assuming $C_{-1} = 0$ and $C_0 = 1$, then the first three coefficients of the expansion (17) can be obtained as

$$C_1 = \frac{Q\eta}{2P}C_0, \qquad C_2 = \frac{1}{2(2P+1)}\left\{-\frac{Q\eta}{2P}\left[(-Q\eta+\tilde{b})-R\right]\right\},$$
$$C_3 = \frac{-1}{6(P+1)}\left\{\frac{-Q\eta+2\tilde{b}}{(2P+1)}\left[R+\frac{Q}{2R}\eta(-Q\eta+\tilde{b})\right]+\frac{Q}{P}\frac{\eta}{2}(R-2)\right\}. \tag{18}$$

At this stage, we can obtain the analytical solution to the radial part of Schrodinger equation. This can be accomplished by breaking the series (17) of the BCH function into Heun's polynomial of degree *n*. Imposing the following conditions on the two coefficients: $C_{n+1} = 0$ and $R = 2n$ with $n = 1,2,3,...$. From the first condition $R = 2n$, it is possible to obtain a formal expression for the energy. Therefore, after adopting this limitation, $R = 2n$, we can simply obtain the energy eigenvalues equation:

$$2\sqrt{\frac{8\mu a}{\hbar^2}+\left(\frac{\mu\omega_c}{\hbar}\right)^2}\left(n+1+\left|m+\frac{\Phi_{AB}}{\Phi_0}\right|\right)$$
$$-\frac{2\mu E_{nm}}{\hbar^2}+\left(m+\frac{\Phi_{AB}}{\Phi_0}\right)\frac{\mu\omega_c}{\hbar}-\frac{\mu^2 b^2}{\hbar^4\left[\frac{8\mu a}{\hbar^2}+\left(\frac{\mu\omega_c}{\hbar}\right)^2\right]}=0, \tag{19}$$

where $n = 1, 2, 3,...$. In fact, the energy equation can be obtained from the necessary conditions of the BCH series, and hence Eq. (13) becomes a polynomial of degree *n* with $\gamma = 2n + \alpha + 2$, [31, 53], too.

Now, to analysis given in Eq. (19), $C_{n+1} = 0$, we consider the cyclotron frequency, $\omega_c$, can be adjusted in such a way that this condition $C_{n+1} = 0$ can be satisfied. This is possible. In this case, we can adjust either the intensity of the magnetic field *B* associated with the uniform volume charge density. In this case, both conditions imposed, $C_{n+1} = 0$ and $R = 2n$, are satisfied and a polynomial solution to the Hune function is obtained. In fact, a result of this analysis is that there exists a relation of $\omega_c$ to the quantum numbers of the system $\{n,l\}$. In this regards, also for more information, see [28, 54-58].

Now, let us calculate the mass spectrum of the heavy quarkonium systems, such as the charmonium and bottomonium made of quark and antiquark of same flavor using the following relation [59]

$$M_{c,b} = 2m_{c,b} + \frac{\hbar^2}{2\mu}\left\{ 2\sqrt{\frac{8\mu a}{\hbar^2} + \left(\frac{\mu\omega_c}{\hbar}\right)^2}\left(n+1+\left|m+\frac{\Phi_{AB}}{\Phi_0}\right|\right) \right.$$
$$\left. +\left(m+\frac{\Phi_{AB}}{\Phi_0}\right)\frac{\mu\omega_c}{\hbar} - \frac{\mu^2 b^2}{\hbar^4\left[\frac{8\mu a}{\hbar^2}+\left(\frac{\mu\omega_c}{\hbar}\right)^2\right]} \right\}, \quad (20)$$

Hence, in Table 1, we show the mass spectra of the charmonium and bottomonium systems. It is obviously seen the large influence of the azimuthal quantum number and magnetic field on the mass spectra of the charmonium and bottomonium systems.

## 3. Thermodynamic properties

Here in this section, we are going to calculate analytically the thermodynamic quantities for a canonical system using the characteristic function. So we begin by calculating the main thermodynamic quantities using the Eq. (19) as follows:

$$\omega_n = \frac{1}{2}\left(\frac{2\mu a}{\hbar^2}+\left(\frac{\mu\omega_c}{2\hbar}\right)^2\right)^{1/2}(2n+\Xi), \quad (21)$$

where

$$\Xi = 2+2\left|m+\frac{\Phi_{AB}}{\Phi_0}\right|+\left(m+\frac{\Phi_{AB}}{\Phi_0}\right)\frac{\mu\omega_c}{\hbar}\left(\frac{2\mu a}{\hbar^2}+\left(\frac{\mu\omega_c}{2\hbar}\right)^2\right)^{-1/2} - \frac{\mu^2 b^2}{\hbar^4}\left(\frac{2\mu a}{\hbar^2}+\left(\frac{\mu\omega_c}{2\hbar}\right)^2\right)^{-3/2}$$

and we have defined $\omega_n = E_n/\hbar$.

At first, we write the characteristic function ($G = \ln Z$ where $Z$ is the canonical partition function [60]) as

$$G = -\sum_{n=1}^{\infty}\ln[1-\exp(-\beta\omega_n)] = -\sum_{n=1}^{\infty}\ln[1-\exp(-2\pi\delta(2n+\Xi))], \quad (22)$$

where $\beta = 1/T$ ($k_B = 1$) is the usual thermodynamic parameter and

$$\delta = \frac{\hbar^2\beta}{\mu}\left(\frac{2\mu a}{\hbar^2}+\left(\frac{\mu\omega_c}{2\hbar}\right)^2\right)^{1/2}\bigg/4\pi.$$

In Appendix A, with the first-order approximation in $-(2\alpha+\Sigma)/2$, the new characteristic function Eq. (22) becomes as follows:

$$G = -\frac{1}{2}(2\alpha+\Sigma)\left[\ln\left(\frac{4\pi}{\beta\Theta}\right) + \frac{\Theta\beta}{2} - \left(2-\frac{\pi^2}{4}\right)\frac{\pi^2}{3\Theta\beta}\right] - \ln\left(\frac{4\pi}{\beta\Theta}\right) - \frac{11\Theta\beta}{48} + \frac{\pi^2}{6\Theta\beta}, \quad (23)$$

where $\Sigma = 1 + \left(m + \dfrac{\Phi_{AB}}{\Phi_0}\right)\dfrac{\hbar\omega_c}{\Theta} - \dfrac{\hbar^2 b^2}{\mu\Theta^3}$ and $\Theta = \dfrac{\hbar^2}{\mu}\left[\dfrac{2\mu a}{\hbar^2} + \left(\dfrac{\mu\omega_c}{2\hbar}\right)^2\right]^{1/2}$

After using the new characteristic function, we can simply calculate the mean energy:

$$U = T^2\frac{\partial G}{\partial T} = -T + \frac{11}{48}\frac{\hbar^2}{\mu}\left[\frac{2\mu a}{\hbar^2} + \left(\frac{\mu\omega_c}{2\hbar}\right)^2\right]^{1/2} + \frac{\mu\pi^2 T^2}{6\hbar^2}\left[\frac{2\mu a}{\hbar^2} + \left(\frac{\mu\omega_c}{2\hbar}\right)^2\right]^{-1/2}$$

$$-\frac{1}{2}(2\alpha+\Sigma)\left\{T - \frac{1}{2}\frac{\hbar^2}{\mu}\left[\frac{2\mu a}{\hbar^2} + \left(\frac{\mu\omega_c}{2\hbar}\right)^2\right]^{1/2} - \frac{\pi^2 T^2\hbar^2}{3\mu}\left[\frac{2\mu a}{\hbar^2} + \left(\frac{\mu\omega_c}{2\hbar}\right)^2\right]^{-1/2}\left(2-\frac{\pi^2}{4}\right)\right\}, \quad (24)$$

where the mean energy at zero temperature $T = 0$ takes on $U = \vartheta(2\alpha+\Sigma+11/12)/4$.

On the other hand, the specific heat can be found as ($C_V = -\partial U/\partial T$)

$$C_V = 1 - \frac{\pi^2\mu T}{3\hbar^2}\left[\frac{2\mu a}{\hbar^2} + \left(\frac{\mu\omega_c}{2\hbar}\right)^2\right]^{-1/2}$$

$$+ \frac{1}{2}(2\alpha+\Sigma)\left\{1 - \frac{2\pi^2\mu T}{3\hbar^2}\left[\frac{2\mu a}{\hbar^2} + \left(\frac{\mu\omega_c}{2\hbar}\right)^2\right]^{-1/2}\left(2-\frac{\pi^2}{4}\right)\right\}, \quad (25)$$

where the specific heat at zero temperature $T=0$ is $C_V = 1 + (2\alpha+\Sigma)/2$.

The free energy ($F = -\ln(Z)/\beta$) is

$$F = \frac{1}{2\beta}(2\alpha+\Sigma)\left[\ln\left(\frac{4\pi}{\beta\Theta}\right) + \frac{\Theta\beta}{2} - \left(2-\frac{\pi^2}{4}\right)\frac{\pi^2}{3\Theta\beta}\right] + \frac{1}{\beta}\ln\left(\frac{4\pi}{\beta\Theta}\right) + \frac{11\Theta}{48} - \frac{\pi^2}{6\Theta\beta^2}, \quad (26)$$

where the free energy at zero temperature $T=0$ is $F = 11\vartheta/48$.

We calculate the entropy $(S = -\partial F/\partial T)$ as

$$S = -\frac{1}{2}(2\alpha+\Sigma)\left[\ln\left(\frac{4\pi}{\beta\Theta}\right) - \left(2-\frac{\pi^2}{4}\right)\frac{2\pi^2}{3\Theta\beta} + 1\right] - \ln\left(\frac{4\pi}{\beta\Theta}\right) + \frac{\pi^2}{3\Theta\beta} - 1. \quad (27)$$

Now, we can obtain the persistent current ($I = -\partial F/\partial\Phi$) [61] as

$$I = -\frac{1}{4\beta}\frac{\omega_c}{\Phi_0}\left[\frac{1}{2\beta}\ln\left(\frac{8\pi}{\beta}\left(\frac{8\mu a}{\hbar^2}+\frac{e^2 B^2}{\hbar^2 c^2}\right)^{-1}\right)\right.$$
$$\left.+\frac{\beta}{4}\left(\frac{8\mu a}{\hbar^2}+\frac{e^2 B^2}{\hbar^2 c^2}\right)-\frac{2\pi^2}{3\beta}\left(\frac{8\mu a}{\hbar^2}+\frac{e^2 B^2}{\hbar^2 c^2}\right)^{-1}\left(2-\frac{\pi^2}{4}\right)\right]. \quad (28)$$

The magnetization ($M = -\partial F/\partial B$) [60, 61] of the present system can be written as

$$M = 2\left[\frac{e^2 B}{c^2}\left(\frac{24b^2\mu^2}{\hbar^6 \tilde{g}^{5/2}}\right)+\frac{e}{2\mu c}\left(m+\frac{\Phi_{AB}}{\Phi_0}\right)\right]$$
$$\times\left[\ln\left(\frac{8\pi\mu}{\hbar^2\sqrt{\tilde{g}}}T\right)+\frac{\hbar^2\sqrt{\tilde{g}}}{4\mu T}-\frac{2\pi^2\mu}{3\hbar^2\sqrt{\tilde{g}}}\left(2-\frac{\pi^2}{4}\right)T\right]T$$
$$-\frac{1}{2}\left[2\left|m+\frac{\Phi_{AB}}{\Phi_0}\right|+1-\frac{8b^2\mu^2}{\hbar^4 \tilde{g}^{3/2}}+\frac{eB}{2\mu c}\left(m+\frac{\Phi_{AB}}{\Phi_0}\right)\right] \quad (29)$$
$$\times\left[\frac{1}{4T\mu}\left(\frac{2\mu a}{\hbar^2}+\frac{e^2 B^2}{c^2}\right)^{-1/2}-\frac{1}{\hbar^6\tilde{g}}+\frac{2\pi^2\mu}{3\hbar^4\tilde{g}^{3/2}}\left(2-\frac{\pi^2}{4}\right)T\right]\frac{e^2 B}{c^2}T$$
$$+\frac{e^2 B}{c^2}\left(\frac{T}{\hbar^6\tilde{g}}-\frac{11}{96\mu\sqrt{\tilde{g}}}-\frac{\pi^2 T^2\mu}{3\hbar^4\tilde{g}^{3/2}}\right),$$

where $\tilde{g} = \frac{8\mu a}{\hbar^2}+\frac{e^2 B^2}{\hbar^2 c^2}$.

## 4. Discussions

For a selected values of model parameters, in Figure 1, we plot the energy states versus the potential parameters $a$ and $b$, it is obvious that energy state increases with the increasing of the values of $a$ whereas decreases with the increasing the values of $b$. For higher states, say $n=2$, the energy is shifted above further.

In Figure 2, we plot the energy versus the magnetic and AB flux fields. It is obvious that the energy increases nonlinearly with the increasing magnetic field for $\Phi_{AB}=2T$, whereas it increases linearly with AB flux field $\Phi_{AB}$ when $B=2T$. In fact, the energy spectrum depends explicitly of the internal AB flux.

In Figure 3, the mean energy $U$ and the specific heat $C_V$ are plotted versus temperature when magnetic field strength $B=1.0T, 1.5T$. It is seen that $C_V$ increases with the increasing $T$ as linear while $U$ decreases with the increasing $T$ as nonlinear. The increase of $C_V$ with $T$ is higher when magnetic field strength is lower, whereas the decrease in $U$ is lower when the magnetic field is lower.

In Figure 4, the free energy $F$ and the entropy $S$ are plotted versus temperature when magnetic field strength $B = 1.0T, 1.5T$. It is noticed that $F$ increases with the increasing $T$ and $S$ decreases with the increasing $T$ as nonlinear. The increase of $F$ with $T$ is higher when magnetic field strength is higher than at $T \geq 0.8K$ with the magnetic field strength increasing the free energy curve becomes lower. The entropy curve is seen higher with strongly applied magnetic field.

In Figure 5, we plot the persistent current $I$ and magnetization $M$ versus the temperature $T$. It is shown that magnetization increases sharply with increasing temperature and the stronger magnetic field has a larger effect on this increase. On the other hand, the persistent current $I$ is decrease linearly slowly with the increasing of temperature. The stronger magnetic field has a lower curve.

In this section, we discuss on our results obtained for the calculated thermodynamic quantities. In the limits when $(2\alpha + \Sigma) << 1$ [Eqs. (23) to (27)], then we can obtain $\Xi = 1$. In this case, the characteristic function of Eq. (23) becomes $G = \ln(\beta\Theta/4\pi) - 11\Theta\beta/48 + \pi^2/6\Theta\beta$. Consequently, we obtain the specific heat: $C_V = 1 - \pi^2/3\beta\Theta$. Further, the Helmholtz free energy: $F = -\pi^2/6\beta^2\Theta + \ln(\Theta\beta/4\pi)/\beta + 11\Theta/48$, the mean energy: $U = -\frac{\partial}{\partial\beta}\ln Z = -\frac{\Theta}{4\pi}\frac{\partial G}{\partial\delta} = \frac{\delta}{\beta}\left[\frac{11\pi}{12} - \frac{1}{\delta} - \frac{\pi}{24\delta^2}\right] = \frac{11\pi\delta}{12\beta} - \frac{1}{\beta} - \frac{\pi}{6\Theta\beta^2}$. Finally, the entropy $(S = -\partial F/\partial T)$ becomes $S = -1 + \ln(\Theta\beta/4\pi) + \pi^2/3\Theta\beta$.

On the other hand, if we study the quantum thermodynamic property for N-particles system such as a Fermi gas or diatomic molecules, we need to expand the partition function as $Z = Z^N$. Further, the dependence on N and volume comes via the dependence on the energy Eigen spectra $E_n$. The energy levels for diatomic molecules can be obtained from the direct solutions of the Schrödinger equation with diatomic molecular potential energy models. Finally, having obtained the energy states, the matter of finding thermodynamic quantities becomes simple and straightforward for the system under consideration [41, 62, 63].

Also, the characteristic function becomes in the form $G = N \ln Z$. Using the new characteristic function, we can simply calculate the other thermodynamic quantities for *N*-particles system such as a Fermi gas and diatomic molecules. In fact, we need to study the behavior of the collective interactions of the charged particles in such systems. Other quantities the dependence on the partition function are simply calculated) similar Eqs. (23)-(28).

## 5. Conclusions

In this work, we extracted a general form of the Schrödinger equation with the Cornell-plus-harmonic potentials under the external magnetic and AB flux fields into curved space. The energy levels and their corresponding wave functions were also obtained under the vector and scalar fields by means of the series method. After finding the energy levels, we obtained mass spectra of heavy quarkonia using the obtained energy levels and we have also calculated the main thermodynamic functions by using the characteristic function for non-relativistic single particle. Also, we discussed on the obtained analytically results. Finally, some results of the thermodynamic quantities are showed, too.

## Conflict of Interests

The author(s) declare(s) that there is no conflict of interest regarding the publication of this paper.

# Appendix A:

A characteristic function is simply the Fourier transform, in probabilistic language. In fact, the characteristic function [64, 65] of a random variable $X$ is defined by

$$X = \int_{-\infty}^{\infty} e^{i\omega x} dx, \tag{A.1}$$

where $i$ is the imaginary unit. The logarithm of $X$ is known as the characteristic function and denoted by $G = \ln X$. In other words, the characteristic function in statistical physics is relevant in relationship with the partition function of an ensemble. In this regard, see a list of some common distributions functions and the corresponding characteristic in Ref. [66]. Recently, Yi and Talkner [67] have determined the work statistics for single particle and $N$-particles by evaluating the characteristic function with the help of the relations in Refs. [68, 69].

At this stage, after expanding the logarithm into Eq. (21), we have

$$\frac{\partial G}{\partial \delta} = -2\pi \sum_{n=1}^{\infty} (2n+\Xi) \sum_{k=1}^{\infty} \exp(-2\pi k \delta (2n+\Xi)). \tag{A.2}$$

We can further write the derivative $G$ by using the complex integral representation [70, 71] $e^{-x} = (1/2\pi i) \int_C ds \, x^{-s} \Gamma(s)$ with $x = 2\pi k \delta (2n+\Xi)$, as

$$\frac{\partial G}{\partial \delta} = -\frac{1}{i} \int_C ds (2\pi \delta)^{-s} \Gamma(s) \zeta(s) \sum_n (2n+\Xi)^{1-s}. \tag{A.3}$$

Now, we can also express Eq. (A.3) in terms of the Euler, Riemann and Riemann's generalized functions in the form

$$\frac{\partial G}{\partial \delta} = -\frac{1}{i} \int_C ds (2\pi \delta)^{-s} \Gamma(s) \zeta(s) 2^{1-s} \zeta\left[s-1, \frac{\Xi}{2}\right]. \tag{A.4}$$

After expanding $\zeta\left[s-1, \frac{\Xi}{2}+1\right]$ up to the third order in $(\Xi-1)$, we get

$$\frac{\partial G}{\partial \delta} = -\frac{\pi}{94\delta^2} \left[\frac{1}{4} - (\pi^2 - 8)(\Xi-1) + (7\zeta[3] - 8)(\Xi-1)^2\right] - \frac{\pi}{12} [3\Xi(\Xi+2) + 2] + \frac{\Xi+1}{2\delta}. \tag{A.5}$$

Here, we applied the Residues theorem for the poles $s - 0,1,2$ in obtaining Eq. (A.4).

**Table 1:** The charmonium, $M_c(GeV)$, and bottomonium, $M_b(GeV)$ mass spectra.

| $\Phi_{AB}(Tesla) = 2T$ $\hbar = c = e = 1$ | | | $b(GeV)$ | $a(GeV)$ | $B(Tesla)$ | $g(GeV)$ | $M_c(GeV)$ | $M_b(GeV)$ |
|---|---|---|---|---|---|---|---|---|
| $n = 1$ $m = 1$ | $m_c(GeV)$ | 1.48 | 0.255 | 0.042 | 2 | 3.068357313 | 20.93557326 | -- |
| | | | | | 4 | | 38.30284487 | -- |
| | $m_b(GeV)$ | 4.68 | 0.465 | 0.143 | 2 | 1.705231169 | -- | 16.12549012 |
| | | | | | 4 | | -- | 21.14492671 |
| $n = 2$ $m = 1$ | $m_c(GeV)$ | 1.48 | 0.255 | 0.042 | 2 | 4.091143084 | 23.72100966 | -- |
| | | | | | 4 | | 43.75008836 | -- |
| | $m_b(GeV)$ | 4.68 | 0.465 | 0.143 | 2 | 2.273641559 | -- | 17.2297556 |
| | | | | | 4 | | -- | 22.99180058 |
| $n = 2$ $m = 2$ | $m_c(GeV)$ | 1.48 | 0.255 | 0.042 | 2 | 5.113928855 | 27.85779742 | -- |
| | | | | | 4 | | 51.90003455 | -- |
| | $m_b(GeV)$ | 4.68 | 0.465 | 0.143 | 2 | 2.842051949 | -- | 18.76137168 |
| | | | | | 4 | | -- | 25.69337531 |

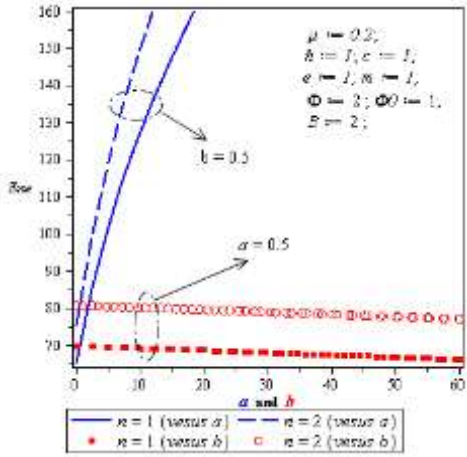

**Fig. 1.** The energy states $E_{nm}$ versus the potential parameter $a$ (blue color online) when $b = 0.5$ and $E_{nm}$ versus the parameter of $b$ (red color online) with $a = 0.5$

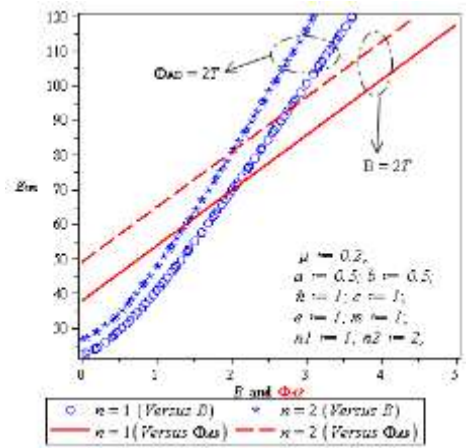

**Fig. 2.** The energy states $E_{nm}$ versus the magnetic field strength $B$ (blue color online) with $\Phi AB = 2.0T$ and $E_{nm}$ versus the AB flux field $\Phi AB$ (red color online) with $B = 2.0T$

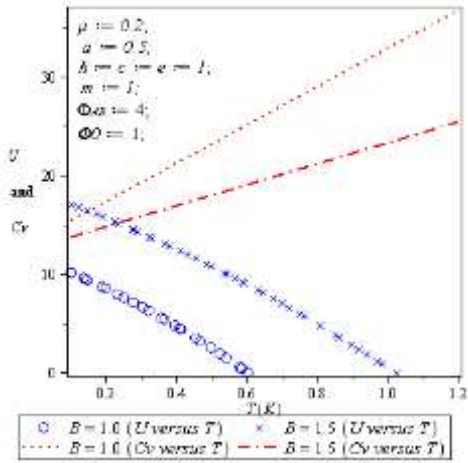

**Fig. 3** The mean energy $U$ versus the temperature $T$ (blue color online) and specific heat $C_v$ versus the temperature $T$ (red color online).

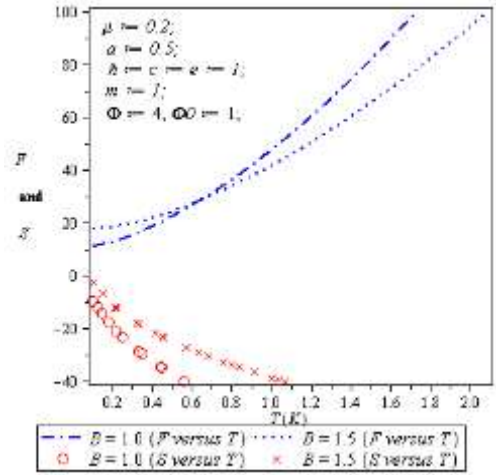

**Fig. 4.** The free energy $F$ versus the temperature $T$ (blue color online) and entropy $S$ versus the temperature $T$ (red color online).

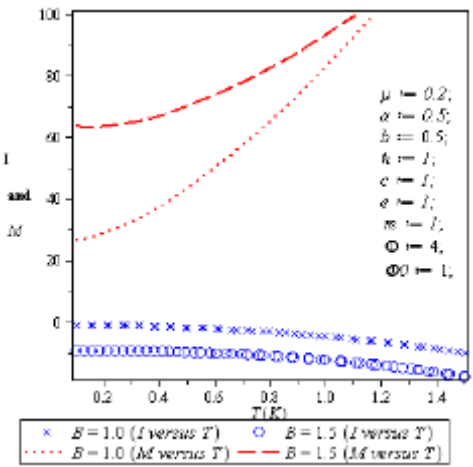

**Fig. 5** The persistent current $I$ versus the temperature $T$ (blue color online) and magnetization $M$ versus the temperature $T$ (red color online).